\documentclass[prb,twocolumn,showpacs]{revtex4}
\usepackage{graphicx}
\usepackage{amsmath}
\usepackage{amssymb}
\usepackage{epstopdf}

\begin{document}
\title{Jastrow-Correlated Wavefunctions for Flat-Band Lattices}
\author{Hao Wang and V. W. Scarola}
\affiliation{Department of Physics, Virginia Tech, Blacksburg,
Virginia 24061, USA}

\begin{abstract}
The electronic band structure of many compounds, e.g., carbon-based
structures, can exhibit essentially no dispersion.  Models of
electrons in flat-band lattices define non-perturbative strongly
correlated problems by default.  We construct a set of
Jastrow-correlated ansatz wavefunctions to capture the low energy
physics of interacting particles in flat bands.  We test the ansatz
in a simple Coulomb model of spinless electrons in a honeycomb
ribbon.  We find that the wavefunction accurately captures the
ground state in a transition from a crystal to a uniform quantum
liquid.
\end{abstract}


\maketitle

\section{Introduction}

Quantum correlated phases and rich phase diagrams can manifest even
in systems with very little contribution from the kinetic energy.
The long-range part of the interaction itself can lead to
interesting and unexpected physics.  Hubbard's work on lattice
models found that the long-range part of the Coulomb interaction
alone supports insulating crystals that stabilize at odd denominator
lattice fillings. \cite{hubbard:1978} His analysis brings up an
interesting question: Can interactions by themselves lead to new and
interesting physics?

Particles hopping with an otherwise large kinetic energy
can interfere to show \emph{no} dispersion in certain lattices
geometries (see, e.g., Ref.~\onlinecite{tasaki:1998a}).  Such flat-bands
arise as particles hop among a few sites and interfere to form
localized states.  At first one would expect such
lattices to exhibit only classical localization effects.  However,
the absence of any dispersion leaves the interaction as the \emph{de
facto} dominant energy scale thus allowing interesting quantum
many-body effects to appear solely from off-diagonal contributions
in the interaction or, equivalently, from non-commuting lattice density operators.

The paucity of theoretical work on interacting flat-band systems
stems from implicit difficulties in solving even the simplest
models. Vanishing band curvature implies very little or no screening
of the interaction.  Off-diagonal terms in the interaction can
therefore be relevant in such systems.  Furthermore, the absence of
dispersion results in a single dominant term in the simplest
flat-band models:
$H_{\text{FB}}=\mathcal{P}^{\dagger}_{\text{FB}}V\mathcal{P}^{\vphantom{\dagger}}_{\text{FB}}$,
where $\mathcal{P}_{\text{FB}}$ projects into a flat band and $V$
is a two-body interaction.  With
no small parameter, conventional perturbative methods fail.  Work on
flat-band lattices has become more pressing with the discovery of
systems hosting flat bands.

The discovery of carbon based nanostructures, e.g., carbon nanotubes
\cite{iijima:1991} and graphene \cite{novoselov:2004}, opened the
possibility of a geometric tuning of electronic band structure.
Detailed calculations show flat bands in a variety of
physically relevant contexts, including: the edges of two
dimensional graphene, \cite{nakada:1996} one dimensional graphene
nanoribbons, \cite{nakada:1996,lin:2009,potasz:2010} hydrogenated graphene
nanoribbons, \cite{kusakabe:2003} collapsed carbon nanotubes,
\cite{lammert:2000} hydrogenated nanotubes, \cite{pei:2006}
graphene dots, \cite{ezawa:2007} and graphene antidots
\cite{vanevic:2009}. Flat bands can also be found in a wealth of
other compounds and engineered in optical lattices. \cite{wu:2007}
The ubiquity of flat-band systems imply that strong interaction effects
demand further study.

We propose that studies in an entirely different context, the
quantum Hall regime, can shed light on the problem of
strong correlation in flat-band lattices.  Models of the quantum
Hall effects project into a single Landau band in the high field
limit.  The degenerate Landau level leaves the Coulomb interaction
as the sole term in the simplest models to yield direct analogies
with models of flat-band lattices in the absence of a magnetic
field.

Parallels between quantum Hall and zero-field lattice formalisms
have been drawn at the Hamiltonian level. Early work
\cite{westerberg:1993,rezayi:1994} compared models of Laughlin's
states \cite{laughlin:1983} in the cylindrical geometry
\cite{thouless:1984} to lattice models of spinless fermions.   Here
it was shown that, on thin cylinders, components of Laughlin's
wavefunction can be thought  of in terms of Hubbard's classical
crystals. \cite{hubbard:1978}  The work on thin cylinders marked a
key advance by noting an implicit one dimensional structure in
quantum Hall models.  Subsequent analyses explored intriguing
aspects of this lattice-quantum Hall connection. \cite{lee:2004}

We argue that a connection between flat-band and quantum Hall
formalisms can be drawn at the level of ansatz wave functions.
Elegant but accurate many-body wavefunctions quantitatively model
the physics of the quantum Hall regime.
\cite{laughlin:1983,jain:1989,dassarma:1997}  Here we build a
similar class of Jastrow-correlated wave functions
\cite{jastrow:1955} as ansatz states for flat-band lattices.  We
numerically test a quasi-one dimensional example in a toy model of a
flat band in graphene nanoribbons (Figure~\ref{graphs}) of up to 476
sites.  We find that the ansatz state accurately captures the ground
state in a remarkable transition from a Wigner crystal to a uniform
quantum liquid driven entirely by interactions.

\section{Jastrow-Correlated Ansatz Wavefunctions}

We construct Jastrow-correlated first
quantized wavefunctions, applicable to flat-band lattices.   We
consider a family of states motivated by Jastrow-correlated states
written in the lowest Landau level. \cite{laughlin:1983,jain:1989}
For a generic two-body interaction that depends only on relative
separation, $H_{\text{FB}}$ commutes with the center of mass
operator.  We may therefore construct states of the form,
$\Psi_{\text{CM}}\Psi_{\text{rel}}$, where $\Psi_{\text{CM}}$ and
$\Psi_{\text{rel}}$ are functions of the center of mass and relative
coordinates, respectively.  We propose that the following set of
unnormalized relative-coordinate wavefunctions are energetically
favorable ansatz states for a broad class of flat-band lattice
models:
\begin{eqnarray}
\vert \Psi_{\text{rel}}^{\nu}\rangle=\prod_{j<k,\{\Lambda\}}
\left(a^{\dagger}_{j,\Lambda}-
a^{\dagger}_{k,\Lambda}\right)^{m_{c}}\vert \psi^{\nu^{*}}\rangle,
\label{psirel}
\end{eqnarray}
where the symbol $\{\Lambda\}$ denotes a set of one dimensional (1D)
chains that cover a bravais lattice.  Each bond-oriented chain
uniquely covers bonds and the two end sites lie at the lattice
edges. The first quantized operator for particle $j$,
$(a^{\dagger}_{j,\Lambda})^{n}$, creates a state localized at the
position ${\bf r}_{n}^{\Lambda}={\bf \delta}_{\Lambda}+n {\bf
b}_{\Lambda}$, where $n\in\mathbb{N}$ and ${\bf \delta}_{\Lambda}$
is the position vector for a unique edge starting site of the chain
$\Lambda$ (e.g., $\delta=0$ for a 1D lattice and
$\{\delta\}=\{(n^{0}_{x},n^{0}_{y}),(n^{0}_{x}\cdot n^{0}_{y}=0)\}$
for a square lattice).  The set of bond vectors ${\bf b}_{\Lambda}$
point from the starting site to a nearest neighbor site along the
chain. (This chain covering procedure must be modified for some
non-bravais lattices.) We define real space single particle basis
states as $\langle {\bf r} \vert (a^{\dagger}_{\Lambda})^{n}\vert
0>=\sqrt{n!} f_{n}(\gamma)w({\bf r}-{\bf r}_{n}^{\Lambda})$, where
$w$ are Wannier functions.  $f$ is a variational function.  For the
model studied here we choose $f_{n}(\gamma)=\exp(\gamma^{2}
n^{2}/2)/\sqrt{n!}$, where $\gamma$ is a variational parameter.

The Jastrow factor in Eq.~(\ref{psirel}) enforces Gutzwiller
projection. It attaches an integer number of correlation holes,
$m_{c}$, to each particle in the wave function $\psi^{\nu^{*}}$ at a
filling, $\nu^{*}$, to thereby form a state at the reduced
filling $\nu=\nu^{*}/(m_{c}\nu^{*}+1)$, akin to a procedure introduced in the quantum
Hall regime. \cite{jain:1989}  Here $\nu$ refers to the number
of particles per basis state. In the following we consider fermions.
The choice $m_{c}=2p$ with antisymmetric $\psi^{\nu^{*}}$ preserves the
antisymmetry of $\Psi_{\text{rel}}$.

In the following we test a 1D example of Eq.~(\ref{psirel}) and a
specific choice for $\psi^{\nu^{*}}$. We consider a Hartree-Fock
state,
$\psi^{\nu^{*}=1}=\text{Det}[(a^{\dagger}_{i})^{n}]\vert0\rangle=\prod_{j<k}
(a^{\dagger}_{j}- a^{\dagger}_{k})\vert0\rangle$.  With this choice
the relative coordinate wavefunction takes a form similar to the
Laughlin state: \cite{laughlin:1983,girvin:1983}
\begin{eqnarray}
\vert \Psi_{\text{1D-rel}}^{\nu}\rangle =
\prod_{j<k}\left(a^{\dagger}_{j}- a^{\dagger}_{k}\right)^{2p+1}\vert
0\rangle.
\label{psi1drelative}
\end{eqnarray}
For an $N$-particle system, the number of particles per basis state
is then $\nu=N/[(2p+1)N-2p]$.  We focus on $p=1$, which
corresponds to $\nu=1/3$ for $N \rightarrow \infty$.

The variational parameter, $\gamma$, tunes the above wavefunctions
between distinct limits.  For $\gamma\rightarrow\infty$ the
wavefunctions describe crystals (Wigner crystals) because the basis
states are highly localized.  For intermediate values,
$\gamma\sim1$, we have a charge density wave.  For
$\gamma\rightarrow0$ we obtain uniform liquid states. To see this
one notes that the Wannier functions can be written in terms of
polynomials in Fourier transform space. \cite{slater:1952}  For a 1D
lattice along $x$ we have $w(x-x_{n})=\mathcal{F}\exp({iq
n})\tilde{w}(q)$, where $\mathcal{F}$ denotes the Fourier integral
over all $q$-space and $\tilde{w}(q)$ is the Fourier transform of
the Wannier function at the edge. Then Eq.~\ref{psi1drelative}
becomes a Laughlin-like state:
\begin{eqnarray}
\lim_{\gamma\rightarrow0}\Psi^{\nu}_{\text{1D-rel}}=\mathcal{F}\prod_{j<k}
\left(e^{iq_j}-e^{iq_k}\right)^{2p+1}\prod_{j}\tilde{w}(q_j),
\end{eqnarray}
which describes a uniform quantum liquid.

\section{Model of Interacting Fermions in a Flat Band}

We test the validity of Eq.~(\ref{psi1drelative}) as an ansatz
ground state for a model of $N$ spinless fermions interacting
through a truncated Coulomb interaction in a flat band of a honeycomb
ribbon.  We consider a second quantized flat-band model:
\begin{eqnarray}
H=\frac{1}{2}\sum_{n_1n_2n_3n_4} V_{\{n\}} \hat{c}_{n_1}^{\dagger} \hat{c}_{n_2}^{\dagger} \hat{c}_{n_3} \hat{c}_{n_4},
\label{Hsecond}
\end{eqnarray}
where the matrix elements $V_{\{n\}}$ are determined by the form of
the interaction $V(|\mathbf{r}-\mathbf{r'}|)$ and flat-band basis
states $\phi$ as:
\begin{eqnarray}
V_{\{n\}}=\int d\mathbf{r} d\mathbf{r'} \phi_{n_1}^{*}(\mathbf{r}) \phi_{n_2}^{*}(\mathbf{r'}) V(|\mathbf{r}-\mathbf{r'}|)\phi_{n_3}(\mathbf{r'})\phi_{n_4}(\mathbf{r}).
\end{eqnarray}
The second quantized operator $\hat{c}_{n}^{\dagger}$ creates a fermion in the state $\phi_{n}$.

Flat-band basis states follow from the lattice structure.  In
the following we work in units of the honeycomb bond length,
$a_{0}=1$, and Coulomb energy, $e^{2}/\epsilon a_{0}$ ($\sim10^{5}
K$ for a carbon bond in vacuum). We consider a quasi-1D
honeycomb ribbon of width $L_{y}=\sqrt{3}N_{y}$ along the $y$
direction. Along the ribbon we allow the range of $x$ to be
infinite, but restrict the centers of local electron orbitals to
lie in $N_{x}$ unit cells along a length of $L_{x}=3N_{x}$. For
simplicity, we assume Gaussians of width $\sim \beta$ localized at
each site of the honeycomb lattice.  The hopping energy between sites can be
used to determine $\beta$ (or vice versa) prior to projection into
the flat-band.  In the honeycomb geometry the flat-band basis states
are:
\begin{eqnarray}
  \phi_{n}(\mathbf{r})=
   N_{\beta}\sum_{s=1}^{2N_{y}} t_{s}
   e^{-(\mathbf{r}-\mathbf{R}_{n}^{s})^2/2\beta^2},& (n=1,..,N_{x}),
  \label{gaussian}
\end{eqnarray}
where $N_{\beta}$ is a normalization constant,
$\mathbf{R}_{n}^{s}=(X_{n}^{s},Y^{s})$ is the location of the $s$th
node in the $n$th unit cell, and the node factor $t_{s}$ has the
form of $t_{2u-1}=t_{2u}=(-1)^{u+1}$ for $u=1,..,N_{y}$.
\cite{lin:2009} The node locations arising from the underlying
honeycomb structure are: $\mathbf{R}_{n}^{s}=(3n-\epsilon_{s},
\sqrt{3}(u-1/2))$, where $\epsilon_{s}=2$ $(\epsilon_{s}=1)$ for
$s=2u-1$ ($s=2u$) and  $u=1,..,N_{y}$. In the following we set
$N_{y}=3$. Using the above Wannier state $\phi_{n}$ as a
single-particle basis state localized at the $n$th unit cell, we can
construct the interaction matrix elements analytically.

We stress that the model considered here idealizes flat-bands of
armchair nanoribbons.  The Gaussian localized states chosen in
Eq.~\ref{gaussian} are only convenient approximations for orbitals
of electrons in flat bands of suspended graphene nanoribbons.  For
example, the $\pi_{z}$ Slater orbitals of carbon atoms could be used
instead. \cite{potasz:2010}  These orbitals are, in contrast,
exponentially localized and loosely correspond to $\beta\sim1$.  We
expect less localization (larger $\beta$) in carbon structures with
adsorbates (e.g., hydrogen \cite{kusakabe:2003}).  It is
straightforward to replace the Gaussian orbitals in our formalism
with density functional theory results for specific systems to make
concrete predictions for experiments.  Furthermore, we have ignored
effects from nearby bands.

We now compute the interaction matrix elements in the flat-band
basis with Gaussian orbitals.  The Fourier expansion of the
interaction is $V(r)=e^{iyQ_0}\int dQ V(Q)e^{iQx}/\pi$ with
$Q_0=\pi/\sqrt{3}$ and
$V(Q)=(e^{2}/\epsilon)\sum_{m}V_{m}L_{m}(Q^{2}+Q_{0}^{2})$, were the
$L_{m}$ are the Laguerre polynomials, $m$ is an odd integer for
spinless fermions, and $V_{m}=\sqrt{\pi}(2m-1)!!/(2^{m+1}m!)$ is a
pseudopotential parameter for the full Coulomb interaction.
\cite{haldane:1983} We choose only  the $m=1$ term, which
corresponds to an effective short-range interaction $V(r)\sim
\nabla^2\delta(\bf{r})$. The matrix elements are then:
\begin{eqnarray}
 V_{\{n\}}=\sum_{\{s\}}
 T_{\{s\}}
 e^{-(r_{n_{1}n_{4}}^2+r_{n_{2}n_{3}}^2+\Delta R^2/2)}\left(\frac{\Delta R^2}{\beta^3}-\frac{\alpha}{\beta}\right),
\label{elements}
\end{eqnarray}
where we define:
\begin{eqnarray}
T_{\{s\}}&=&V_1T_{s_{1}s_{4}}T_{s_{2}s_{3}}^{*}/(\sqrt{2\pi}N_{\beta}^{4}),\nonumber \\
T_{ss'}&=&t_{s}t_{s'}\int_{0}^{L_{y}}dy e^{[{iQ_{0}y}-((y-Y^{s})^2+(y-Y^{s'} )^2)/(2\beta^2)]},\nonumber \\
r_{n_{1}n_{4}}&=&(X_{n_{1}}^{s_1}-X_{n_{4}}^{s_4})/(2\beta),\nonumber \\
r_{n_{2}n_{3}}&=&(X_{n_{2}}^{s_2}-X_{n_{3}}^{s_3})/(2\beta),\nonumber \\
\Delta R&=&(X_{n_{1}}^{s_1}+X_{n_{4}}^{s_4}
-X_{n_{2}}^{s_2}-X_{n_{3}}^{s_3})/(2\beta), \text{and} \nonumber\\
\alpha&=&\beta^{-2}+Q_{0}^2.
\label{elements2}
\end{eqnarray}

\section{Numerical Test of Ansatz}

We test our ansatz by solving Eqs.~(\ref{Hsecond})-(\ref{elements2})
with numerical diagonalization on finite-sized ribbons.  We restrict
our ground state to the Hilbert space with the center of mass
at the center of the ribbon.  In the thermodynamic
limit the center of mass is a good quantum number but in finite
sized systems we omit edge effects in extreme unit cells by
restricting our basis. We test several different system sizes and
show below that our results have converged with increasing system
size.

\begin{figure}[t]
\centerline{\includegraphics [width=3.0 in] {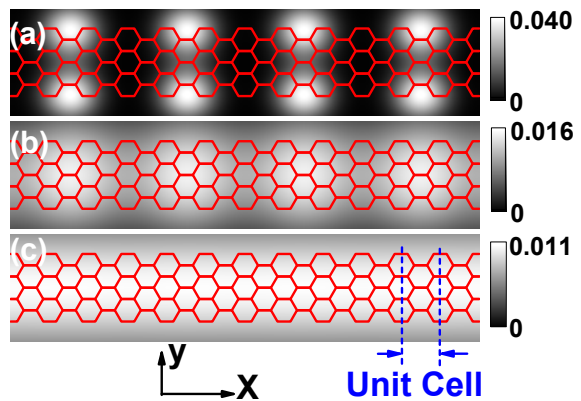}} \caption{(Color
online) Greyscale plot of numerically calculated solutions for the
electron density of the flat-band Coulomb problem in a honeycomb
ribbon with $N=12$ electrons and $N_{x}=34$ unit cells zoomed into
12 central unit cells. Sites sit at vertices in the depiction of the
underlying lattice. The three graphs show three distinct states that
arise from increasing width parameter of basis states: (a) Wigner crystal
($\beta=2.0$), (b) Charge density wave ($\beta=3.5$), and (c)
Uniform state ($\beta=5.0$).} \label{graphs}
\end{figure}

\begin{figure}[t]
\centerline{\includegraphics [width=3 in] {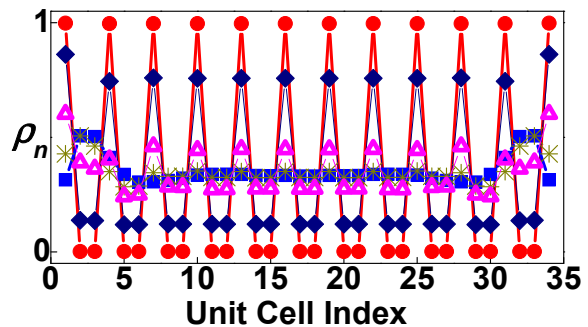}} \caption{(Color
online) The local density of each unit cell, $\rho_{n}$, plotted
versus the unit cell index along the honeycomb ribbon with $N=12$ particles.
The circles, diamonds, triangles, stars, and squares are obtained
from width parameters $\beta=2, 3, 4, 5,$ and $6$,
respectively. The lines are a guide to the eye.} \label{density}
\end{figure}

Figure~\ref{graphs} plots a zoom in of the quasi-1D charge density
over a few central unit cells of an $N=12$ and $N_x=34$ system
computed from our flat-band model,
Eqs.~(\ref{Hsecond})-(\ref{elements2}). The charge density is
$\rho(\textbf{r})=\sum_{n=1}^{N_x}|\phi_{n}(\textbf{r})|^2\rho_{n}$,
where
$\rho_{n}=<\Psi_{\text{FB}}|\hat{c}_{n}^{\dagger}\hat{c}_{n}^{\vphantom{\dagger}}|\Psi_{\text{FB}}>$
is the local density of the $n$th unit cell and $\Psi_{\text{FB}}$
is the exact ground state.  The density is plotted against the
underlying honeycomb lattice to show that each unit cell consists of
14 sites in this example. Figure~\ref{graphs} shows the evolution
from a Wigner crystal configuration (a), to a charge density wave
(b), and then to a uniform state (c) as we increase the extent of
basis states. The same features are exhibited in
Figure~\ref{density}, the local density plot for each unit cell.

\begin{figure}[t]
\centerline{\includegraphics [width=3 in] {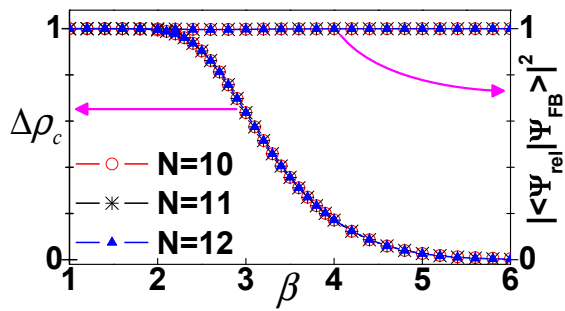}} \caption{(Color
online) The left axis plots the deviation in local density between
neighboring unit cells at the sample center versus the basis
state width parameter for three different system sizes .  The
right axis plots the overlap between the ansatz
state, Eq.~(\ref{psi1drelative}), and the numerically computed ground
state of Eqs.~(\ref{Hsecond})-(\ref{elements2}).} \label{overlap}
\end{figure}

We check the validity of Eq.~(\ref{psi1drelative}) in describing the
ground state of Eqs.~(\ref{Hsecond})-(\ref{elements2}) by taking
overlaps.  We generate Eq.~(\ref{psi1drelative}) numerically by
noting that it is the exact eigenstate of a model of the lowest
Landau level of electrons on a cylinder.
\cite{thouless:1984,rezayi:1994} We obtain the maximum overlap with
the choice $\gamma=3/\beta$.  The right axis of Figure~\ref{overlap}
plots the overlap of Eq.~(\ref{psi1drelative}) with the ground state
of Eqs.~(\ref{Hsecond})-(\ref{elements2}). The left axis plots the
deviation in the local density from the center to the nearest
neighboring unit cell,
$\Delta\rho_{\text{c}}\equiv|\rho_{(N_x/2)}-\rho_{(N_x/2-1)}|$.
$\Delta\rho_{\text{c}}=1$ and 0 correspond for a crystal and a
uniform state, respectively.  We find that the overlap is
essentially constant and larger than $99.61\%$ for all $\beta$ thus
showing that the ansatz wavefunction accurately captures the ground
state of the truncated Coulomb interaction in \emph{all} parameter
regimes.  We expect the overlap to decrease as we increase the range
of the interaction or change the form of the Wannier function.
Overlaps can be improved by choosing a better variational form for
$f(\gamma)$.

It is interesting to note that the ansatz state adiabatically
connects the Wigner crystal to a uniform, Jastrow-correlated liquid
state based on the Laughlin wavefunction.  This crossover was
pointed out in the quantum Hall context in Ref.~\onlinecite{rezayi:1994}.
Here we have shown that a mapping of the quantum Hall state to a
zero-field flat-band lattice preserves this crossover.

\begin{figure}[t]
\centerline{\includegraphics [width=3 in] {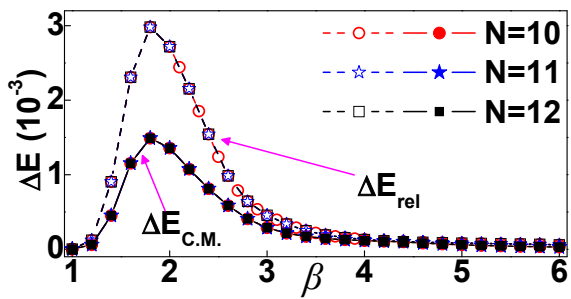}} \caption{(Color
online) The excitation energy plotted versus the basis state width parameter
for three different system sizes.
Open (closed) symbols indicate the energy gap in the relative (center of mass) degrees of freedom.}
\label{gap}
\end{figure}

Figure~\ref{gap} plots the energy gap as a function of $\beta$.  In
an infinitely large system, excitations can either leave the center
of mass intact (excitations in the relative coordinate degrees of
freedom) or shift the center of mass (center of mass excitations).
We find a small but finite gap for the uniform state (large $\beta$)
in the relative degrees of freedom.  The gap appears to remain
robust against system size suggesting that the gap we have computed
is indeed a bulk gap, thus signaling the intriguing possibility of
an incompressible uniform state.  We note, however, that the center
of mass gap becomes comparable to the relative coordinate gap for
large $\beta$.  It is possible that excitations in the relative
degrees of freedom are spatially large in this case and therefore
sensitive to edge effects. Periodic systems will allow a more
accurate estimate of the bulk gap.

\section{Summary}

We have constructed a general class of Jastrow-correlated
wavefunctions applicable to flat-band lattices.  We studied a toy
model of spinless fermions in a flat-band in graphene nanoribbons in
the absence of a magnetic field. We find that the wavefunction
accurately captures an intriguing transition from a Wigner crystal
to a Jastrow correlated liquid, driven entirely by interactions. Our
method can be generalized to accurately model graphene nanoribbons
and other flat-band lattice structures
\cite{nakada:1996,lin:2009,kusakabe:2003,lammert:2000,pei:2006,ezawa:2007,vanevic:2009,potasz:2010,wu:2007}
and can be adapted to study magnetism recently explored in
experiments on graphene nanoribbons. \cite{tao:2011}

The Jastrow-correlated wavefunctions constructed here are versatile.
They can be combined with density functional theory estimates of
Wannier functions to simultaneously capture band structure and the
effects of strong interactions.   They can be generalized to apply
to a wide variety of lattices, describe correlated bosons, or
incorporate spin.  For example $\Psi_{\text{rel}}$ can quantify a
connection between Gutzwiller-RVB and quantum Hall states.
\cite{jain:2009}  Gutzwiller projected $d$-wave spin-singlet paired
states in a two dimensional square lattice can be constructed by
writing $\psi^{\nu^{*}}$ as an antisymmetric product of singlet
paired states, $\propto\mathcal{A}[g({\bf r}_{1}-{\bf r}_{2})g({\bf
r}_{3}-{\bf r}_{4})... ]$.  Here an effective pairing among
$\psi^{\nu*}$ particles could arise from an over-screening of long
range repulsion, as in the quantum Hall regime,
\cite{dassarma:1997,read:2000,scarola:2000} but in flat or
narrow-band lattice models.

\section{Acknowledgments}

We thank K. Park for helpful comments and the Thomas F. Jeffress and
Kate Miller Jeffress Memorial Trust, Grant No. J-992, for support.

\end{document}